\documentclass[twocolumn,PRD,showpacs,nofootinbib,amsmath,amssymb,superscriptaddress]{revtex4}

\pdfoutput=1

\usepackage{graphicx}
\usepackage{dcolumn}
\usepackage{bm}
\usepackage{hyperref}
\usepackage{color}
\usepackage{rotate}
\usepackage{fancyhdr} 
\usepackage{cancel}
\usepackage{multirow}
\usepackage{ulem}

\usepackage{aas_macros}

\begin{document}

\title{The Cosmic Expansion History from Line-Intensity Mapping}
\author{Jos\'e Luis Bernal}
\affiliation{ICC, University of Barcelona, IEEC-UB, Mart\' i i Franqu\` es 1, E08028 Barcelona, Spain}
\affiliation{Dept.\ de F\' isica Qu\` antica i Astrof\' isica, Universitat de Barcelona, Mart\' i i Franqu\` es 1, E08028 Barcelona, Spain}
\affiliation{Institut d'Estudis Espacials de Catalunya (IEEC), E08034 Barcelona, Spain}
\author{Patrick C.~Breysse}
\affiliation{Canadian Institute for Theoretical Astrophysics, University of Toronto, 60 St.\ George Street, Toronto, ON, M5S 3H8, Canada}
\author{Ely D.~Kovetz}
\affiliation{Department of Physics, Ben-Gurion University, Be'er Sheva 84105, Israel}

\begin{abstract}
Line-intensity mapping (LIM) of emission {from} star-forming galaxies can be used to measure the baryon acoustic oscillation (BAO) scale as far back as the epoch of reionization. This provides a standard cosmic ruler to constrain the expansion rate of the Universe at redshifts which cannot be directly probed otherwise. 
In light of  growing tension between measurements of the current expansion rate using the local distance ladder and those inferred from the cosmic microwave background, extending the constraints on the expansion history to bridge between the late and early Universe is  of paramount importance. 
Using a newly derived methodology to robustly extract cosmological information from LIM, which minimizes the inherent degeneracy with unknown astrophysics, we show that present and future experiments can gradually improve the measurement precision of the expansion rate history, ultimately reaching percent-level constraints on the BAO scale.
Specifically, we provide detailed forecasts for the SPHEREx satellite, which will target the H$\alpha$ and Lyman-$\alpha$ lines, {for a near-future stage-2 experiment targeting CII}, and for the ground-based COMAP instrument---as well as a future stage-3 experiment---that will target the CO rotational lines. 
Besides weighing in on the so-called Hubble tension, reliable LIM cosmic rulers can enable wide-ranging tests  of dark matter, dark energy and modified gravity.
\end{abstract} 

\maketitle

Measurements of its expansion history have always been at the heart of our understanding of the Universe. This dates back to Hubble's ninety-year-old discovery of the expansion~\cite{1929PNAS...15..168H}, which marked the dawn of scientific cosmology, to the discovery just  twenty years ago that the expansion is currently accelerating~\cite{Riess_DE98,Perlmutter_DE99}, which provided the last piece in what is now widely considered to be the standard model of cosmology, known as $\Lambda$CDM.

To this date, there are two main observational handles on the expansion rate of the Universe. One relies on calibrating a distance ladder towards standard-candle type-Ia supernovae  in our {\it local} cosmic neighborhood. The other comes from the {\it distant} early Universe, via the extraction of the baryon acoustic oscillation (BAO) scale, which is robustly determined by the sound horizon at recombination, from  cosmic microwave background (CMB) measurements. These opposite handles anchor the so-called direct and inverse distance ladder, respectively~\cite{Cuesta:2014asa}. 

One of the most pressing mysteries in cosmology today is why these measurements have been growing in tension~\cite{Freedman:2017yms}. 
The tension between the value of $H_0$  as directly measured in the local Universe~\citep{RiessH0_19} and the value inferred from Planck CMB measurements~\citep{Planck18_pars} (when $\Lambda$CDM is assumed), has surpassed $4\sigma$ in statistical significance~\cite{RiessH0_19,Wong:2019kwg}.
If real, the discrepancy can be re-framed as a mismatch between the two anchors of the cosmic distance ladder~\cite{BernalH0}. Low redshift observations~\cite{Alam_bossdr12,Scolnic_pantheon} strongly constrain the expansion rate of the Universe under the $\Lambda$CDM prediction. The difference could be due to modifications in the very early Universe or in the epoch $3\lesssim z\lesssim 1000$, which is too faint to probe with discrete galaxy surveys.

In this {\it Letter}, we explore how upcoming and future line-intensity mapping (LIM)~\cite{Kovetz_IMstatus}  experiments can bridge the gap between the local and distant observations to enable the measurement of the full expansion history of the Universe~\cite{Karkare:2018sar,Kashilinsky_LyaBAO,Kashlinsky_LyaBAO_review}. 
LIM experiments can achieve this feat as they integrate light from all emitting sources, including the numerous galaxies at high redshift which are  too faint to be detected individually. Thus, by targeting bright, easily-identifiable spectral lines and covering a wide range of frequencies with high spectral resolution, LIM can provide tomographic maps of specific line emission across extended epochs in the history of the Universe. 

However, a key challenge to the extraction of precise cosmological information from line-intensity maps is the degeneracy with astrophysics, as this line emission is determined by star-formation processes and by both intra and intergalactic gas physics. Unfortunately, these processes are likely never to be understood with the required precision for cosmological analyses. To circumvent this predicament, we will follow a novel methodology---presented in detail in a companion paper~\cite{Bernal_IM}--- to isolate the cosmological information and optimize its extraction. {{Encouragingly, we find that next-decade experiments can reach $
\sim\!1\%$ constraints on the expansion history over redshifts $2\lesssim z\lesssim 6$, and within $4\%$-$10\%$ error at $z\lesssim 8$, providing (robust) invaluable insight on the Hubble tension.}}

We start by reviewing how the BAO scale can be constrained via measurements of the LIM power spectrum.
When transforming redshifts and sky positions into distances to compute a power spectrum, a fiducial cosmology must be assumed. However, the inferred distances will be distorted if the adopted cosmology does not match the correct one~\cite{Alcock_Paczynski}. This effect---named after Alcock and Paczynski---is the key to interpreting the measured BAO scale in a cosmological framework as a standard ruler. 

In Fourier space,  the relation between the true and measured wavenumbers along the line of sight and in the transverse directions is given by $k^{\rm tr}_\parallel=k^{\rm meas}_\parallel/\alpha_\parallel$ and $k^{\rm tr}_\perp=k^{\rm meas}_\perp/\alpha_\perp$, where the rescaling parameters are defined as:
\begin{equation}
\alpha_\perp = \frac{D_A(z)/r_{\rm s}}{\left( D_A(z)/r_{\rm s}\right)^{\rm fid}},\qquad \alpha_\parallel = \frac{\left(H(z)r_{\rm s}\right)^{\rm fid}}{H(z)r_{\rm s}},
\label{eq:scaling}
\end{equation}
where $D_A(z)$ and $H(z)$ are the angular diameter distance and the Hubble expansion rate at redshift $z$, respectively, $r_{\rm s}$ is the sound horizon at radiation drag, and fiducial quantities are denoted by `fid'. 
 An external prior on $r_{\rm s}$ can  then be applied to infer absolute measurements of the  expansion history.

Line-intensity fluctuations provide a biased tracer of the underlying density field (which contains the cosmological information), but are also strongly linked to astrophysical processes. 
To efficiently separate between the astrophysical and cosmological dependences in the LIM power spectrum, 
it is useful to group all the degenerate parameters (see Ref.~\cite{Bernal_IM} for details) and express it as:
\begin{equation}
P(k,\mu) = \left(\frac{\langle T\rangle b\sigma_8  + \langle T\rangle f\sigma_8 \mu^2 }{1+0.5\left( k\mu\sigma_{\rm RSD}  \right)^2}\right)^2\frac{P_m(k)}{\sigma_8^2} + P_{\rm shot},
\label{eq:paramPk}
\end{equation}
where $\mu= \hat{k}\cdot\hat{k}_\parallel$. 
 The power spectrum depends on $\langle T\rangle$, the expected value of the brightness temperature and the luminosity bias $b$ {(both proportional to the first moment of the line's luminosity function, weighted by the halo bias in the case of $b$)},  the growth rate $f$,  the amplitude $\sigma_8$ and the matter power spectrum $P_m(k)$. $P_{\rm shot}$ {(proportional to the second moment of the luminosity function)} is the shot-noise power spectrum, and $\sigma_{\rm RSD}$ accounts for  redshift-space distortions on small scales.
Note that we drop the $z$-dependence from all quantities to simplify notation. From Eq.~\eqref{eq:paramPk}, the set of parameter combinations that can be directly measured 
 at each independent redshift bin and observed patch of sky is $\vec{\theta} = \lbrace  \alpha_{\perp}, \alpha_\parallel, \langle T\rangle f\sigma_8, \langle T\rangle b\sigma_8, \sigma_{\rm RSD}, P_{\rm shot}\rbrace$. 

The multipoles of the observed  power spectrum are~\cite{Bernal_IM}
\begin{equation}
\begin{split}
\tilde{P}_{\ell}(k^{\rm meas}) = & \frac{H(z)}{H^{\rm fid}(z)}
\left(\frac{D_A^{\rm fid}(z)}{D_A(z)}\right)^2\frac{2\ell+1}{2}  \times \\ 
& \times \int_{-1}^{1}{\rm d}\mu^{\rm meas} \tilde{P}(k^{\rm true},\mu^{\rm true})\mathcal{L}_{\ell}(\mu^{\rm meas}),
\end{split}
\label{eq:multipole_scale}
\end{equation}
where $\mathcal{L}_{\ell}$ is the Legendre polynomial of degree $\ell$ and  the observed power spectrum is determined by the observational window function, using $\tilde{P}(k,\mu) =  W(k,\mu)P(k,\mu)$.

 In this work we focus on spectral lines related to star formation, which are brighter than HI with respect to the corresponding foregrounds, and hence may be more promising to provide higher signal-to-noise measurements 
  as far back as the epoch of reionization. 
We consider Lyman-$\alpha$, H$\alpha$, {CII,} and CO(1-0), and  follow the models and prescriptions of~Refs.~\cite{Silva_Lyalpha}, \cite{Gong_lines}, \cite{Silva_CII} and \cite{Li_CO_16}, respectively. The standard approach in these models is to use a set of scaling relations (calibrated from simulations and/or dedicated observations) to associate a star-formation rate to a halo of a given mass, and then relate it to the line luminosity. This can then be integrated over the halo mass function to get an expected signal \cite{Lidz:2011dx,Breysse:2014uia,Breysse2019}. In our analysis below we adopt the fiducial parameter values presented in these works,  naively interpolating (or extrapolating) them to other redshifts, as needed. Naturally, herein lies the largest uncertainty in  our forecasts. However, we emphasize that our use of Eqs.~\eqref{eq:paramPk},\eqref{eq:multipole_scale} ensures that the influence of the astrophysical uncertainties is minimal. While in a pessimistic scenario the amplitude of the signal may be lower than we forecast, the marginalization over the astrophysical parameters renders the analysis less susceptible to {incorrect modelling of the astrophysics}
\footnote{{Only in some cases, such as regarding the radiative transfer of the Lyman-$\alpha$ line, where the observed emission extends beyond the dark matter halo, this procedure might not be general enough and fail to marginalize over these effects. We leave the study of this scenario for future research.}}.

To estimate the potential of LIM BAO measurements, we forecast  measurements of the following planned and future experiments targeting the spectral lines above: 

1.\ SPHEREx~\citep{spherex}: {Launching in 2023, besides performing a wide galaxy survey, SPHEREx will carry out LIM surveys targeting Lyman-$\alpha$ and H$\alpha$ emission, as well as H$\beta$ and the oxygen lines OII and OIII (albeit  with lower significance, making them less suitable for cosmological analyses).} 
 Specifically, we consider the SPHEREx deep survey, which will cover $200\,{\rm deg}^2$ of the sky with higher sensitivity than the all-sky survey. 
 SPHEREx will have $6.2\,{\rm arcsec}$  angular resolution at full-width half maximum, and $R=\nu_{\rm obs}/\delta_\nu=41.4$  {and 150} spectral resolution at $0.75<\lambda_{\rm obs}<4.1$ 
and  $4.1<\lambda_{\rm obs}<4.8$ $\mu$m{, respectively}  (where $\delta\nu$ is the width of the frequency channel, and $\nu_{\rm obs}$  and $\lambda_{\rm obs}$ are the observed frequency and wavelength, respectively) \cite{spherex}. We restrict ourselves to $0.75<\lambda_{\rm obs}<4.1$ $\mu$m, as that is the H$\alpha$ wavelength range modelled  in Ref.~\cite{Gong_lines}.  We assume a sensitivity\footnote{Olivier Dor\'e, private communication.} such that the product $\nu_{\rm obs}\sigma_N/\sqrt{t_{\rm pix}}=$ $\lbrace 3.04,1.49,0.81,0.61\rbrace$ nW/m$^2$/sr/pixel for H$\alpha$ at $\nu_{\rm obs}=\lbrace  29.5,15.7,10.9,8.3\rbrace\times 10^4$GHz; and $\lbrace 3.59, 3.15, 2.54 \rbrace$ nW/m$^2$/sr/pixel for Lyman-$\alpha$ at $\nu_{\rm obs}=\lbrace  36.6, 30.8, 25.2\rbrace\times 10^4$GHz. 

{2.\ CII-StageII~\citep{Silva_CII,Lidz_APforeground}:  Conceived as an upgrade to CONCERTO~\cite{Concerto}, this is a baseline for a near-future experiment 
to ensure detection of the CII power spectrum in the case this line is dimmer than expected. It assumes coverage of the 200-300 GHz   frequency band with a Noise Equivalent Flux density of 5 mJy$\sqrt{\rm s}$, probing $100\,{\rm deg}^2$ of the sky with 64 detectors during 2000 hours and angular and spectral resolutions of 0.5 arcmin and 0.4 GHz.}

3.\ COMAP~\citep{Cleary_COMAP}: Already observing, COMAP is a single-dish, ground-based telescope targeting the CO lines in the frequency band $26$-$34$ GHz. 
 The instrument houses 19 single-polarization detectors with an angular resolution of $4\,{\rm arcmin}$ and channel width of $\delta\nu = 15.6\,{\rm MHz}$. The expected system temperature is $T_{\rm sys}\sim 40\,{\rm  K}$. 
The first phase, COMAP1, will observe 2.25 deg$^2$ of the sky for $t_{\rm obs}=6000$ hours with one telescope, while the second phase, COMAP2 will observe for $10000\,{\rm hours}$ using four additional telescopes (for a total of 95 detectors), all with an improved spectral resolution of $\delta\nu=8\,{\rm MHz}$.  
We assume a sky coverage of $\Omega_{\rm field} = 60$ deg$^2$ for COMAP2, which optimizes the significance of the power spectrum measurement.


4.\ IMS3 (CO): Finally, we envision a next generation (or stage 3) of LIM experiments---IMS3, in short. We conceive a ground-based CO experiment as an upgrade of COMAP, 
and assume it will integrate over the frequency range $12$-$36\,{\rm GHz}$ for 10000 hours using 1000 detectors over 1000 deg$^2$ of the sky, with spectral and angular resolution of 2 MHz and 4 arcmin, respectively. 
Based on Refs.~\cite{Prestage_GBT,Murphy_nextVLA}, we assume $T_{\rm sys} = {\rm max}\left[20,\,\nu_{\rm obs}\left({\rm K}/{\rm GHz}\right)\right]$.

To track the evolution of the signal {and probe the expansion of the Universe at different times}, we divide the observed volumes into redshift bins. We consider in each case non-overlapping, independent redshift bins such as $\log_{10}\left[\Delta(1+z)\right] = \log_{10}\left[\Delta(\nu/\nu_{\rm obs})\right] = 0.1 $ (where $\nu$ is the rest frame frequency), with the corresponding effective redshift located in the center of the frequency bin. This results in four ({three}) bins for the SPHEREx H$\alpha$ {(Lyman-$\alpha$)} observations{, two bins for CII-StageII, and one and five redshift bins for COMAP and IMS3 CO, respectively.} 

\begin{table*}[]
\vspace{0.2cm}
\centering
\resizebox{\textwidth}{!}{
\begin{tabular}{|c||c|c|c|c||c|c|c||c|c||c||c||c|c|c|c|c|}
\hline
 & \multicolumn{4}{c||}{SPHx (H$\alpha$)} & \multicolumn{3}{c||}{SPHx (Ly$\alpha$)} & \multicolumn{2}{c||}{CII-StII} & COMAP1 & COMAP2 & \multicolumn{5}{c|}{IMS3 (CO)}  \\ \hline
$z$ & 0.55 & 1.90 & 3.20 & 4.52 & 5.74 & 7.01 & 8.78 & 5.91 & 7.44 & 2.84 & 2.84 & 2.73 & 4.01 & 5.30 & 6.58 & 7.87  \\ \hline\hline
$\sigma_{\rm rel}\left(D_A(z)/r_{\rm s}\right) \%$ & 5.1 & 2.9 & 3.0 & 4.1 & 3.0 & 6.8 & 19.2 & 6.4 & 34.8 & 10.8 & 3.2 & 0.7 & 0.6 & 0.6 & 0.9 & 1.3  \\ \hline
$\sigma_{\rm rel}\left(H(z)r_{\rm s}\right) \%$ & 44.3 & 22.0 & 23.1 & 30.2 & 26.6 & 43.4 & 91.9 & 9.2 & 50.1 & 13.6 & 3.9 & 0.8 & 0.8 & 0.8 & 1.2 & 1.7  \\ \hline
\end{tabular}%
}
\caption{Forecasted 68\% confidence-level marginalized relative constraints on $D_A(z)/r_{\rm s}$ and $H(z)r_{\rm s}$ from SPHEREx (H$\alpha$, Lyman-$\alpha$), {CII-StageII}, and COMAP1 and COMAP2  and IMS3 (CO) observations (expressed in percentages). }
\label{tab:BAO_const}
\end{table*}

Following Ref.~\cite{Bernal_IM}, we apply the Fisher matrix formalism~\citep{Fisher:1935,Tegmark_fisher97} to forecast constraints on the  BAO measurements using the LIM power spectrum multipoles up to the hexadecapole. 
We  take the fiducial values for the $\Lambda$CDM model parameters from the best fit to the combination of the full CMB Planck dataset and BAO from SDSS galaxies~\cite{Planck18_pars,Alam_bossdr12}. Finally, we use the halo mass function and halo bias fitting function introduced in Ref.~\cite{Tinker_hmf2010}. 

\begin{figure}
\centering
\includegraphics[width=\linewidth]{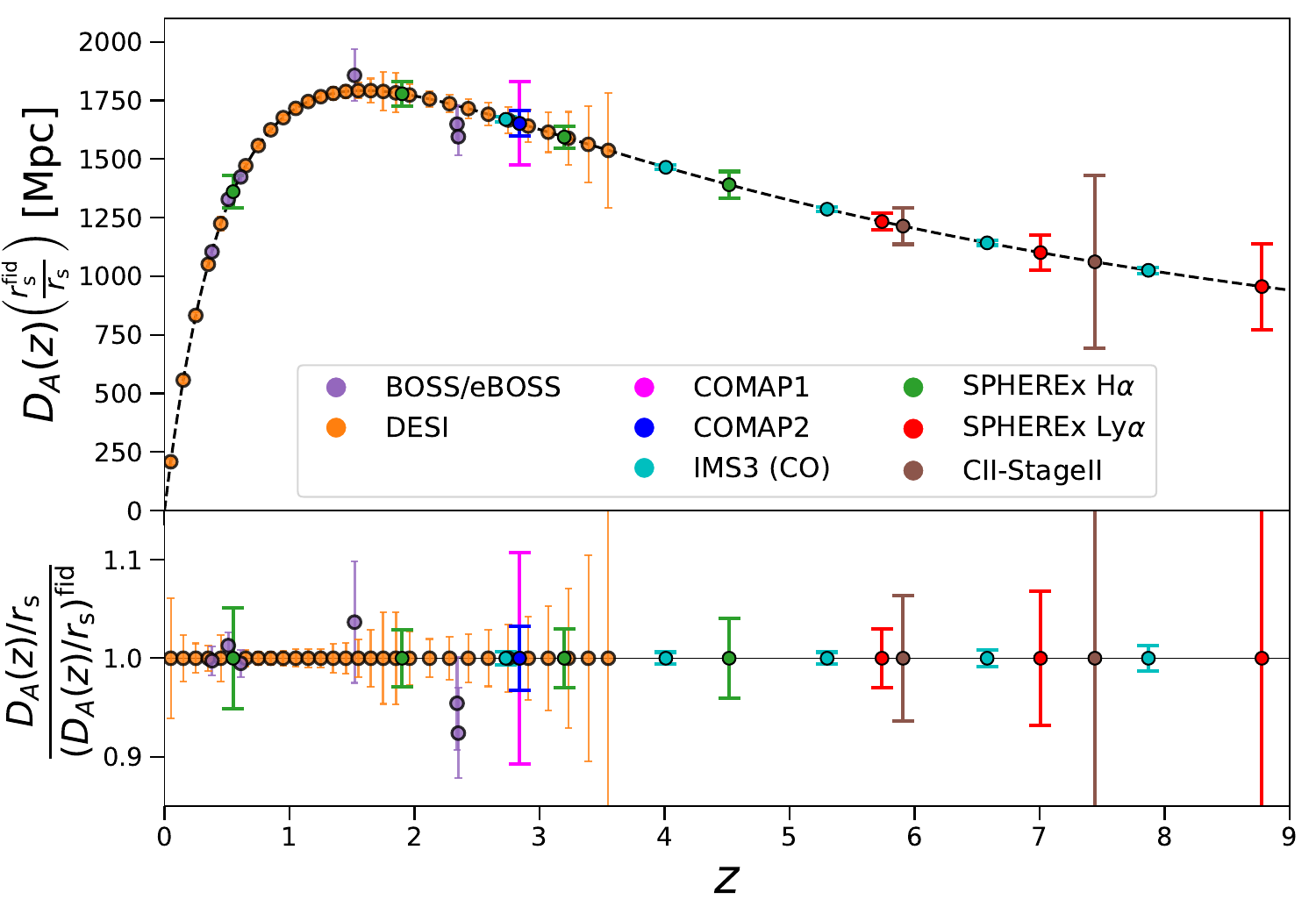}
\includegraphics[width=\linewidth]{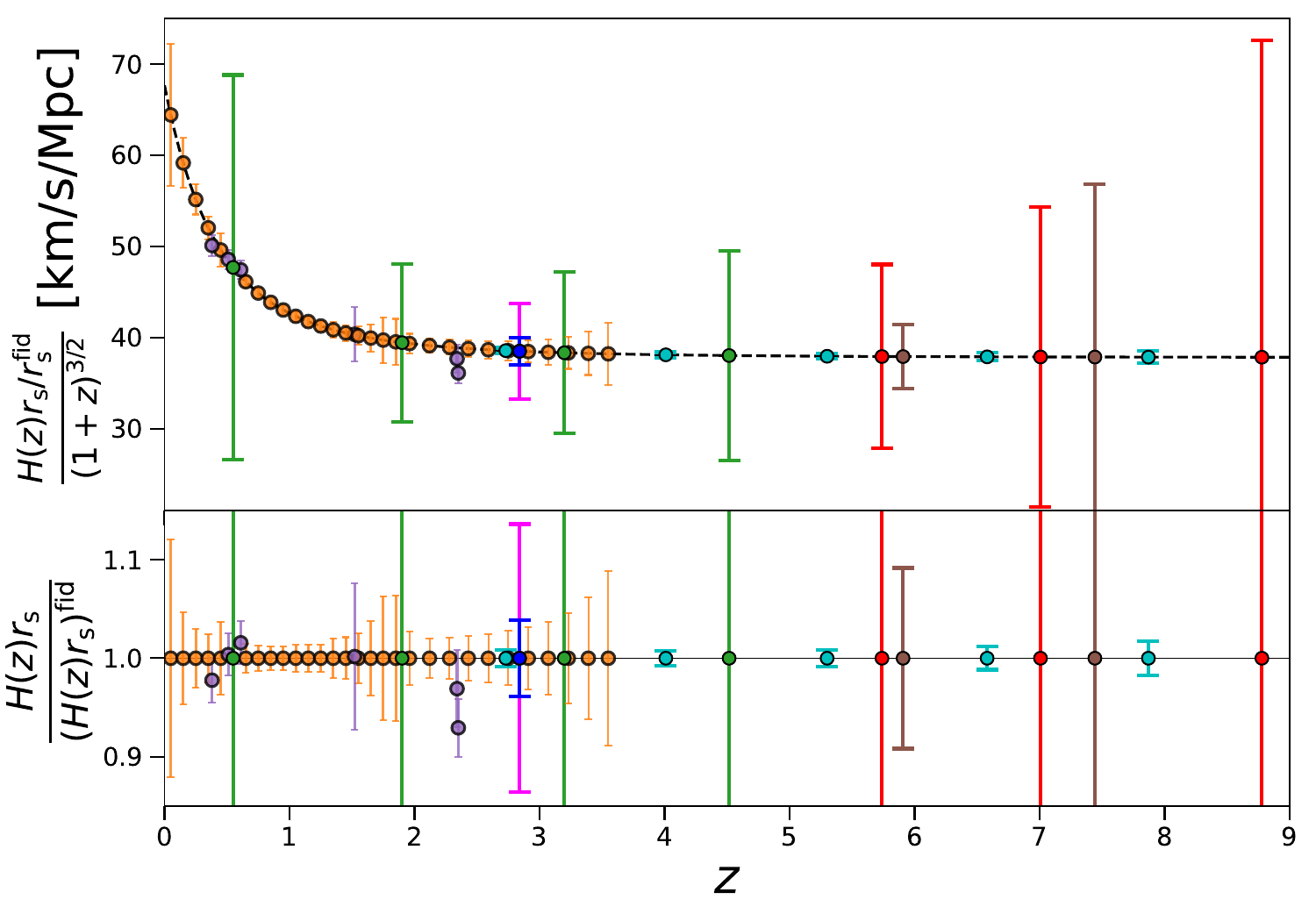}
\caption{ 68\% confidence-level marginalized current and forecasted constraints on the angular diameter distance (top) and Hubble expansion rate over $(1+z)^{3/2}$ (bottom) as a function of redshift, weighted by the ratio between the actual sound horizon at radiation drag and its fiducial value. Estimated constraints from LIM observations of H$\alpha$ (green) and Lyman-$\alpha$ (red) lines using SPHEREx, {of CII using CII-StageII}, and of CO using COMAP1 (pink), COMAP2 (blue) and IMS3 (cyan) are
compared with existing and upcoming measurements from galaxy surveys.}
\label{fig:constraints}
\end{figure}

In table~\ref{tab:BAO_const}, we report forecasted marginalized 68\% confidence-level relative constraints on $D_A(z)/r_{\rm s}$ and $H(z)r_{\rm s}$ from the surveys considered. In Fig.~\ref{fig:constraints}, we compare them with existing~\cite{Alam_bossdr12,GilMarin_qsoeBOSS,eBOSS_Lyalpha_auto,eBOSS_Lyalpha_cross} and prospective~\cite{desi} measurements from galaxy surveys. Due to its poor spectral resolution, SPHEREx constraints on $H(z)r_{\rm s}$ are expected to be very weak{, but it will provide $\sim 3\% \!-\! 7\%$ precision constraints on $D_A(z)/r_{\rm s}$ almost up to $z=7$. }
 This is not the case for COMAP {or CII-StageII}, whose power to constrain $D_A/r_{\rm s}$ and $H(z)r_{\rm s}$ is more balanced. While COMAP1 will be less precise, both COMAP2, with a precision of {$\sim 3\% - 4\%$}, SPHEREx (only in the transverse direction), {and CII-StageII} will be competitive with existing measurements, and not fall much behind of DESI~\cite{desi}. 

Meanwhile, a future IMS3 (CO) experiment would yield percent-level precision at {$2.7\lesssim z\lesssim 7.8$}.  The next generation of LIM experiments will thus allow a precise determination of the expansion rate of the Universe up to the epoch of reionization. The first redshift bin would overlap with BAO measurements from the Lyman-$\alpha$ forest observed with galaxy surveys, allowing a calibration of the LIM BAO.  {{Note that the Lyman-$\alpha$ Forest is inherently more sensitive to the radial direction, while the opposite is often true for LIM experiments.}}

 \begin{figure}
\centering
\includegraphics[width=\linewidth]{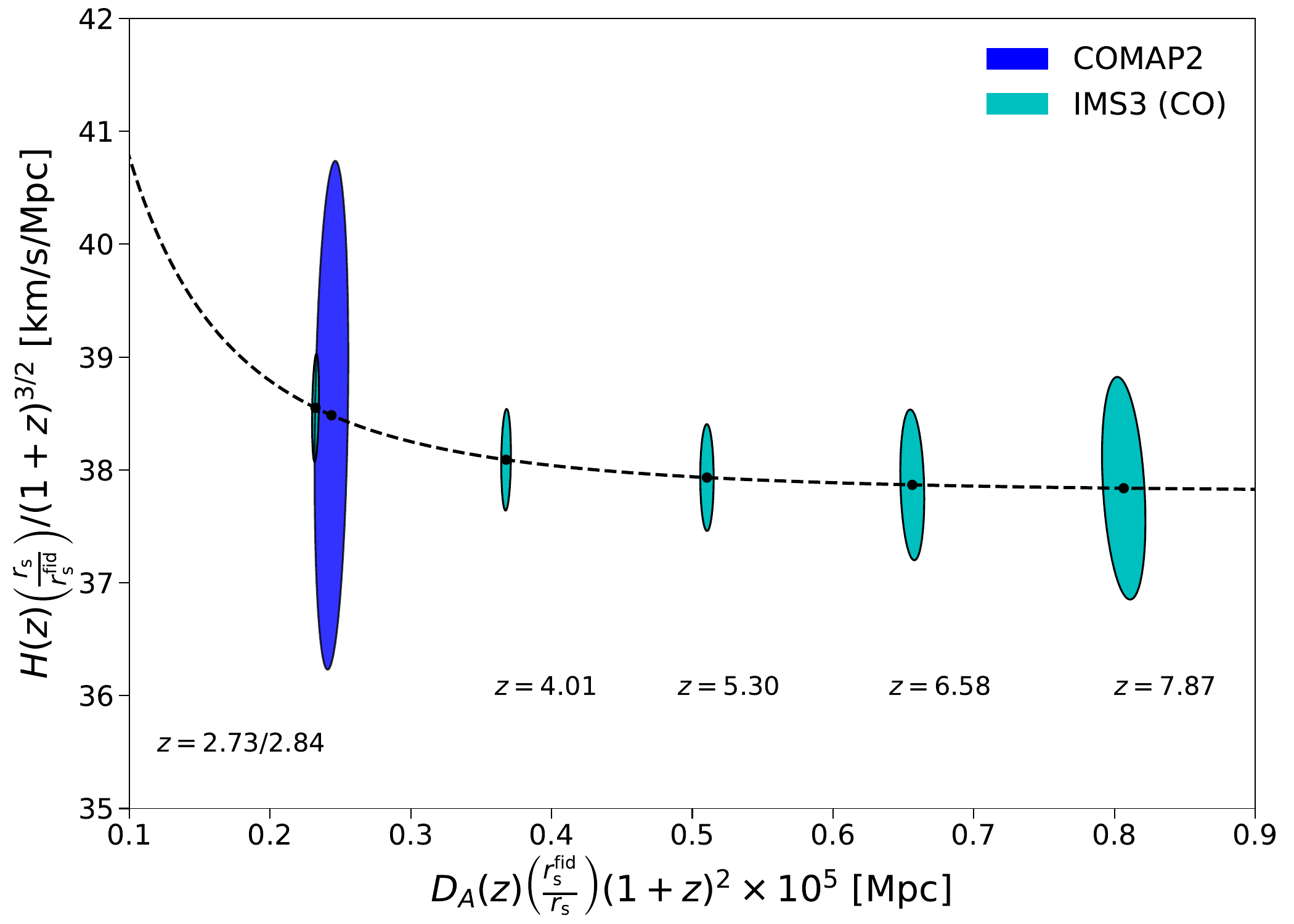}
\caption{ 68\% confidence-level marginalized forecasted constraints on the plane $D_A(1+z)^2$ - $H/(1+z)^{3/2}$, when both are weighted by the ratio between the actual sound horizon at radiation drag and its fiducial value. We show forecasts for the CO line using COMAP2 (blue) and IMS3 (cyan). 
 }
\label{fig:constraints2D}
\end{figure}

In Figure~\ref{fig:constraints2D}, we show the forecasted marginalized constraints on the plane $D_A(1+z)^2$ - $H/(1+z)^{3/2}$, for the case of CO.  The correlation between radial and transverse BAO measurements changes with redshift, although this is mainly driven by the transformation of angular and radial resolutions into physical distances.
The  correlations between cosmological parameters contrained using BAO measurements, such as the Hubble constant $H_0$ and the matter density parameter today $\Omega_M$  also change with redshift (see e.g.\ Ref.~\cite{Cuceu_BAOH0} for results from galaxy surveys).

It should be emphasized that although the evolution of $H(z)$ at $3\lesssim z\lesssim 10$ is completely determined by $\Omega_M$ under $\Lambda$CDM, we still lack direct empirical evidence of the expansion history in these epochs (CMB lensing provides an integrated {constraint}, roughly peaked at $z\sim2$~\cite{Aghanim:2018oex}). There are models that predict other behaviors, including modified gravity theories or dark matter decaying into lighter dark particles (see e.g.\ Refs.~\cite{Raveri_DEMG,Kyriakos_dcdm}). These models, as well as alternative modifications of the cosmic expansion in the matter dominated era, could potentially reduce the Hubble tension. Still, model-independent reconstructions of the Hubble parameter currently remain unconstrained beyond $z\sim 0.7$~\citep{BernalH0,StandardQuantities,Joudaki:2017zhq}.

Therefore, perhaps the most convincing way to demonstrate the potential of LIM BAO measurements is to consider a model-independent expansion history. Following Ref.~\cite{BernalH0}, we parametrize $H(z)$ with natural cubic splines. We locate the nodes of the splines at $z=\lbrace 0.0,0.2,0.6,1.6,2.4,4.0,6.5,9.0\rbrace$ {(chosen to optimize the constraining power)} and fit the values of $H(z)$ using uniform priors and the following data\footnote{We run MonteCarlo Markov Chains using \texttt{emcee}~\cite{emcee}, publicly available at \url{dfm.io/emcee/}.}: the local measurement of $H_0$~\cite{RiessH0_19};  Type-Ia supernovae  (SNeIa)~\cite{Scolnic_pantheon}; BAO from galaxies~\cite{Beutler11,Ross15,Alam_bossdr12,Kazin14_wz,eBOSS_baoz72}, quasars~\cite{GilMarin_qsoeBOSS} and the Lyman-$\alpha$ forest~\cite{eBOSS_Lyalpha_auto,eBOSS_Lyalpha_cross}; and $r_{\rm s}$ measured from 2018 Planck  data~\cite{Planck18_pars}. In addition, we {include mimicked $D_A(z)$ and $H(z)$ measurements from LIM BAO observations by SPHEREx, CII-StageII and  IMS3 (CO). We assume the covariance matrix given by the corresponding Fisher matrix, and draw the central values from such covariance matrices centered on the fiducial values.} 

 \begin{figure}
\centering
\includegraphics[width=\linewidth]{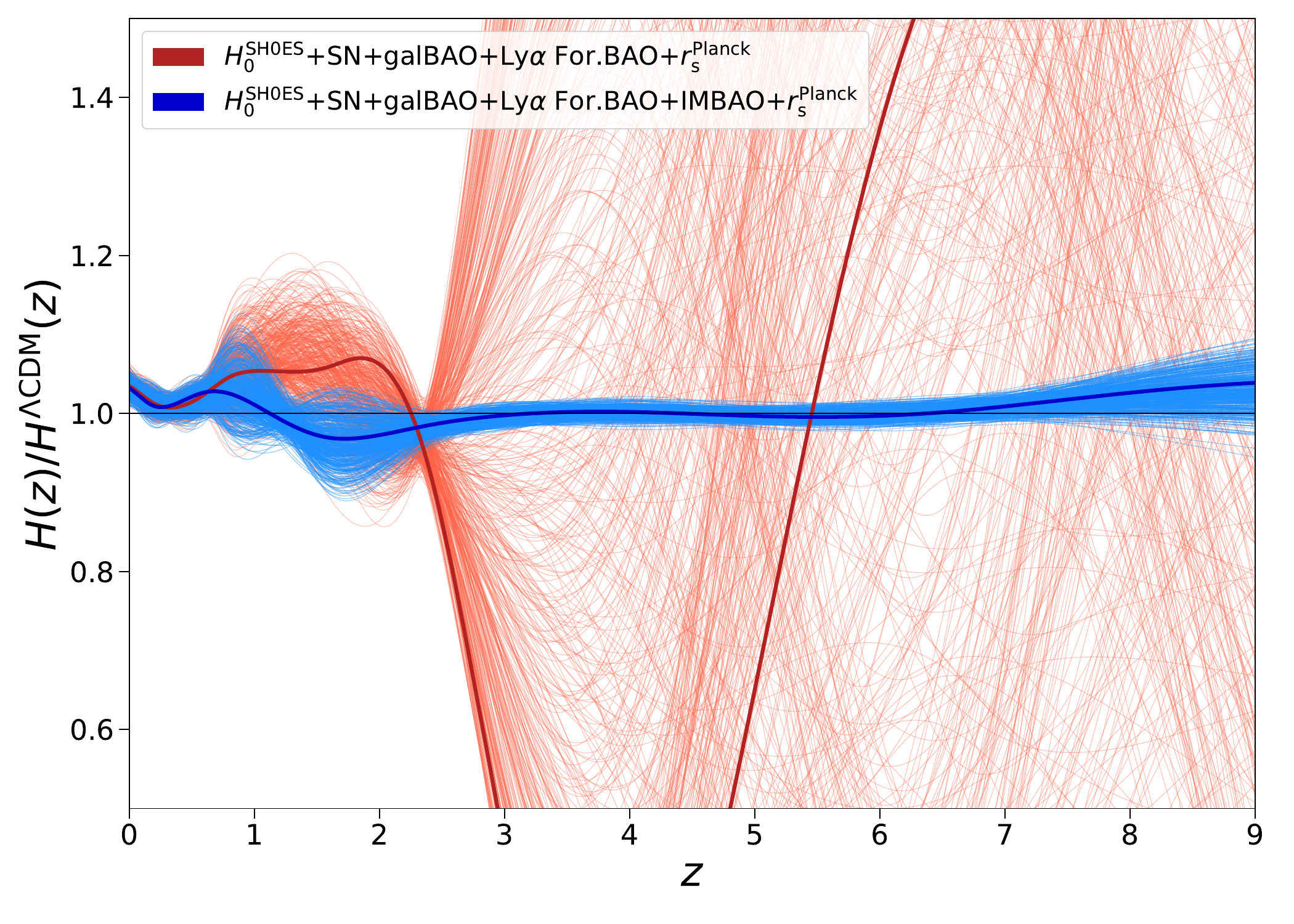}
\caption{Constraints on the model independent reconstruction of $H(z)$ using existing data (red) and including the LIM BAO (blue). We show the best fit with wide solid lines, and 500 random samples drawn from the 68\% confidence-level region using thin solid lines.}
\label{fig:splines}
\end{figure}

In Fig.~\ref{fig:splines}, we show the constraints on the model-independent reconstruction of $H(z)$, with and without the estimated LIM BAO measurements. While the $H_0$ and $r_{\rm s}$ measurements constrain the amplitude of $H(z)$, the shape is constrained by the SNeIa and BAO data. We can see how, without the LIM BAO measurements, the  expansion  history is completely unconstrained beyond $z\sim 2.4$, given the lack of observations. Fortunately, LIM BAO measurements will enable us to fill this gap and extend the constraints on the expansion history of the Universe up to $z\sim 9$, {with constraints below the $\sim 2\%-3\%$ level for $2\lesssim z\lesssim 7$, and at $\sim 4\%$ level at $z\sim 9$}. 

Results using our methodology as forecasted in this {\it Letter} would represent an incredible achievement for cosmology and provide an unique way to directly measure how the Universe expands at $z\sim~3 - 5$ and up to $z\sim7-9$. 
 In the future, standard sirens~\citep{standard_siren,Feeney:2018mkj,Chen:2017rfc} might also achieve this goal, although the  sensitivity needed {{to obtain precise measurements of neutron star mergers and their electromagnetic counterparts at these redshifts}} is considerably more {demanding} than what is expected from upcoming and future experiments. 
 Moreover, also using LIM observations, measurements of the velocity-induced acoustic oscillations~\cite{Munoz_vao_meth} at cosmic dawn, as proposed in Ref.~\cite{Munoz_vao}, can potentially constrain the expansion rate at $15\lesssim z\lesssim 20$. Thus, thanks to LIM experiments, our ignorance about the expansion of the Universe may be limited to $20\lesssim z\lesssim 1000$, where there is little room to accommodate a solution to the $H_0$ tension. LIM can thus shed light on various potential scenarios suggested to solve the  tension~\cite{Karwal:2016vyq,Poulin_EDE,Stephon_H0dilaton,Lin:2019qug,Kreisch_selfnu,Yang:2019nhz,Archidiacono:2019wdp,Hooper:2019gtx,Pandey:2019plg,Pan:2019gop,Desmond:2019ygn,Agrawal:2019lmo,Vagnozzi:2019ezj}.

 Foregrounds and line-interlopers  might  degrade the results reported here, but {it is expected that they do not become an insurmountable limitation for spectral lines related to star formation.} 	
  Moreover, in the coming years there will be several LIM observations which will overlap with galaxy surveys. Cross-correlations between different tracers will make it possible to subtract this contamination from the LIM signals (see e.g., Refs.~\cite{Silva_Lyalpha,Sun_foregrounds}). Finally, we must emphasize that the luminosity functions of spectral lines at high redshift are still highly uncertain. Although using our methodology to disentangle between astrophysical and cosmological dependences, this should not bias the measurements~\cite{Bernal_IM},   it may certainly affect their precision  by modifying the signal-to-noise ratio if the amplitude of the LIM power spectrum turns out to be lower than assumed here. We have accounted for this by choosing line emission models whose predictions are neither too optimistic nor too conservative.

To conclude, LIM experiments can provide precise and robust measurements of the BAO scale up to the epoch of reionization at $z\lesssim 9$. These observations will provide superb constraints on the expansion history of the Universe, probe  models of exotic dark matter, dynamical dark energy, modified gravity, etc., and in general open a new discovery space in the high-redshift universe.

We acknowledge useful conversations with Tzu-Ching Chang, Olivier Dor\'e, H\'{e}ctor Gil-Mar\'{i}n and Licia Verde. JLB is supported by the Spanish MINECO under grant BES-2015-071307, co-funded by the ESF.

\bibliography{biblio}
\bibliographystyle{utcaps}

\end{document}